\documentclass[a4,11pt]{article}
 \usepackage{amsmath,amssymb,color,graphics,amscd,amsfonts,epsf}
 \usepackage[dvipdfm]{graphicx}
 \usepackage{mathrsfs}
 \newcommand{\vev}[1]{\left\langle #1 \right\rangle}
 \newcommand{\diag}{\mathop{\rm diag}}
 \allowdisplaybreaks[4]
 
 \date{}
 
\begin{document}
 {\usefont{T1}{pcr}{m}{n}}
 \DeclareFontShape{T1}{pcr}{mc}{n}{<-> pcrr8tn}{}
 \begin{flushright}
 KUNS-2115 \\
 \end{flushright}
 
 \vspace{0.1cm}

 \begin{center}
 \textbf{\LARGE 
Non-equilibrium thermodynamics \\
near the horizon and holography}
\end{center}
\vspace{0.1cm}
\begin{center}
Mitsutoshi Fujita \footnote
{
E-mail address : mfujita@gauge.scphys.kyoto-u.ac.jp}

\vspace{0.3cm}

{\it Department of Physics, Kyoto University,
Kyoto 606-8502, Japan}\\
\vspace{0.1cm}

\end{center}

\vspace{1.5cm}

\begin{center}
{\bf Abstract}
\end{center}
$\usefont{T1}{pcr}{mc}{n}$
Small perturbations of a black brane are interpreted as small deviations from thermodynamic equilibrium in a dual theory with the AdS/CFT correspondence. In this paper, we calculate hydrodynamics of the dual Yang-Mills theory in the gravity side using membrane paradigm. This method is different from the usual AdS/CFT correspondence and evaluate classical solutions not at boundaries but at a place slightly away from a horizon. There are sound modes or shear modes for gravity perturbation. For sound modes, such calculation at the horizon has not yet been done. Then, we find that boundary stress tensor at the horizon satisfies conservation law in flat space and can represent dissipative parts of stress tensor in the dual theory by holography. Using them, we can read off directly shear and bulk viscosity of the dual theory.
Quasi-normal modes are solutions to linearized equations obeyed by classical fluctuations of a gravitational background subject to specific boundary conditions and are also gauge-invariant quantities. We use solutions for each fluctuation that compose such quantities and show that quasi-normal modes are consistent with the membrane paradigm.  

\newpage

\section{Introduction}
The AdS/CFT correspondence~\cite{Ma,Gu,Wi} makes it possible to analyze quantum states of 
some field theories by obtaining classical solutions of the corresponding supergravity on the AdS spacetime (in the zero temperature case) or a black brane background with an event horizon (in the finite temperature case). In the real-time finite-temperature AdS/CFT correspondence, boundary conditions at the horizon and the infinity must be imposed on classical solutions on such a background. On the horizon, we choose the incoming wave boundary condition. 

One of the interesting applications of the finite temperature AdS/CFT correspondence is the calculation, in the gravity side, of the viscosity coefficients of Yang-Mills theories.  
There are many methods to calculate viscosity coefficients using the AdS/CFT correspondence. 
First, the shear viscosity in $\mathcal{N}=4$ supersymmetric Yang-Mills theory is calculated using real-time finite-temperature AdS/CFT correspondence~\cite{Policastro:2001yc}. This method evaluates classical 
solutions of the supergravity at the boundary of the spacetime. 
 
Second, the viscosities in Yang-Mills theories are calculated by considering the quasi-normal modes on the supergravity background. As we will see later, quasi-normal modes are invariant under infinitesimal diffeomorphisms, and they vanish at the infinity and satisfy the incoming boundary conditions on the horizon. 
We identify the quasi-normal frequencies with the singularities of the retarded Green's function in Yang-Mills theories.
By using further the constitutive relations of the dual Yang-Mills theories, we can read off the shear and the bulk viscosities. This method is more suited than the first one for evaluating the   
subleading terms of the dispersion relation.
 
Third, the membrane paradigm~\cite{So,Ko2} is a more direct method to calculate the viscosity coefficients. In this method, we define a conserved quantity on stretched horizons which is composed of $U(1)$ gauge fields. By imposing the above boundary conditions at the infinity and on the horizon, we can reconstruct a part of the hydrodynamic equation of the Yang-Mills theories.  From this, we can read off the shear viscosity of the Yang-Mills theories. However, the membrane paradigm has a fewer applications than the above two methods. In particular, the bulk viscosity has not been calculated in the membrane paradigm approach.       
 
 The main purpose of this paper is to read off the bulk viscosity from the coefficients of the hydrodynamic equation obtained by the membrane paradigm. In this analysis, it is crucial to start with the quasi-local stress tensor of Brown and York~\cite{Br} defined not at the boundary but near the horizon. (See \cite{Ha} for an analysis of Brown and York's quasi-local stress tensor in the AdS/CFT correspondence.) Then, we deform the quasi-local stress tensor so that it satisfies the conservation law in flat spacetime and can represent the dissipative part of the stress tensor in the dual theory by holography. Using this deformed quasi-local stress tensor, we can read off the shear and the bulk viscosity of the Yang-Mills theories.

The organization of the rest of this paper is as follows.
 In section 2, we review the quasi-normal modes in near-horizon geometries of black p-branes
~\cite{Ko,Mas} and give solutions for each fluctuation that composes quasi-normal modes. We use the quasi-normal modes to count the degrees of freedom of the system. 
In section 3, we review the Brown and York's boundary stress tensor and the constitutive relations of the stress tensors. Following the membrane paradigm approach, we construct a conserved stress tensor near the horizon from which the viscosity can be read off. In other words, we calculate directly the shear viscosity, the bulk viscosity and the sound velocity from the quasi-normal mode using the membrane paradigm. Our result means that the quasi-normal mode analysis in the AdS/CFT correspondence is consistent with the membrane paradigm approach. 
The final section (section 4) is devoted to a summary and discussions.

\section{Quasi-normal modes and the solutions of
fluctuations $H_{\mu\nu}(r)$}
In this section, we first review (in section 2.1) on the quasi-normal modes of the gauge invariant fluctuations on near horizon geometries of black $p$-branes \cite{Ko,Mas}. However, the gauge invariant fluctuations are insufficient for our analysis using the membrane paradigm. Therefore, in section 2.2, we present solutions for $H_{\mu\nu}$ which constitute the gauge invariant fluctuations. 
\subsection{Quasi-normal modes: review}
The quasi-normal modes are solutions to the linearized equation for fluctuations with imaginary frequency.
They are invariant under infinitesimal diffeomorphisms, vanish at the infinity and satisfy the incoming boundary conditions on the horizon. 
The quasi-normal modes are necessary for analyzing to quasi-local stress tensor near the horizon.

Let us use following action in $p+2$ dimensions which is obtained from the supergravity action 
by dimensional reduction \cite{Boo} and consists of the metric and a dilaton:
 \begin{equation} 
I_{\rm bulk}=\dfrac{1}{16\pi G_{p+2}}\int _{\mathcal{M} } d^{p+2}x\sqrt{-g}
 \left(R(g)-\dfrac{\beta}{2}\partial _{\mu}\phi \partial ^{\mu}\phi 
-\mathcal{P}(\phi)\right) -\dfrac{1}{8\pi G_{p+2}}\int _{\partial \mathcal{M}
}d^{p+1}x \sqrt{-\gamma}\Theta ^{\mu}_{\ \mu},  \label{ibulk}
\end{equation}
with
\begin{equation}
\dfrac{1}{G_{p+2}}=\dfrac{2\pi ^{\frac{9-p}{2}}L^{8-p}}{\Gamma
 (\frac{9-p}{2})G_{10}},\quad \mathcal{P}(\phi )=\dfrac{(7-p)(p-9)}{2L^2}e^{\frac{4(3-p)}{p(7-p)}\phi (r)},\quad \beta =\dfrac{8(9-p)}{p(7-p)^2}. 
 \label{comps}
\end{equation} 
 In (\ref{ibulk}) and (\ref{comps}), $\gamma_{\mu\nu}$ is the induced metric on the boundary, 
 $\Theta^{\mu\nu}$ is the extrinsic curvature, and $L$ is the radius of dimensional reduction.
The following black $p$-brane geometry is a solution to the equation of motion of (\ref{ibulk}):  
\begin{alignat}{4} 
&ds_{p+2}= -c_{T}^2(r)dt^2+ c_X^2(r)\sum ^p_{i=1}dx_i^2+
c_R^2(r)dr^2,  \notag \\
&c_{T}^2(r)=\left(\dfrac{r}{L}\right)^{\frac{9-p}{p}}f(r),\quad c_X^2(r)
=\left(\dfrac{r}{L}\right) ^{\frac{9-p}{p}},\quad c_R^2(r)=\dfrac{1}{f(r)}
\left(\dfrac{r}{L}\right) ^{\frac{p^2-8p+9}{p}}, \label{geometry}
\end{alignat} 
 and
 \begin{equation} 
\phi (r)=-\dfrac{(3-p)(7-p)}{4}\ln \left(\dfrac{r}{L}\right),
 \end{equation} 
with $f(r)=1-(r_0/r)^{7-p}$.

 Let us consider fluctuations of the bulk fields around the black $p$-brane solution.
 We focus on a single Fourier component that propagates along the coordinate $z=x^p$:
  \begin{alignat}{4} 
&\delta g_{\mu\nu}(t,z,r)=e^{-i(\omega t-qz)}h_{\mu\nu}(r), \notag \\
&\delta \phi (t,z,r)=e^{-i(\omega t-qz)}\varphi (r).
\label{f5}
 \end{alignat} 
 In the following, we omit writing the exponential factor $e^{-i(\omega t-qz)}$.
 Then, the $SO(p-1)$ symmetry of the space $(x_1,x_2,..,x_{p-1})$ groups the fluctuations into three channels:
\begin{alignat}{4} 
&\text{scalar} \to \text{sound channel}:h_{tt}, h_{tz}, h_{zz}, h_{rr}, h_{tr},
h_{zr}, h, \varphi \notag \\
&\text{vector} \to \text{shear channel}:h_{ta}, h_{za}, h_{ra} \notag \\
&\text{traceless tensor} \to \text{scalar channel}:h_{ab}-\delta _{ab}\dfrac{h}{p-1} \label{3channel}
 \end{alignat}
 with $a,b=1,...,p-1$ and $h =\sum _ah_{aa}$. Instead of $h_{\mu\nu}$, we use the following $H_{\mu\nu}$,
\begin{alignat}{4} 
&h_{tt}(r)=c_T^2(r)H_{tt}(r),\quad h_{\mu j}(r)=c_X^2(r)H_{\mu \alpha}(r),
\end{alignat} 
 with $\alpha =1, .., p$. We impose the gauge fixing condition $\delta g_{\mu r}=0$. As already mentioned above, quasi-normal modes \cite{Ko} are gauge invariant  fluctuations under the residual  gauge transformations $x^{\mu}\to x^{\mu}+\xi ^{\mu}$, $\delta g_{\mu\nu}\to \delta g_{\mu\nu}-\nabla _{\mu}\xi _{\nu}-\nabla _{\nu}\xi _{\mu}$ and $\delta\phi\to\delta \phi +\partial ^{\mu}\phi\xi _{\mu}$, and are given by
\begin{alignat}{4} 
&\begin{cases} Z_0=q^2\dfrac{c_{T}^2}{c_X^2}H_{tt}+2q\omega H_{tz}+\omega ^2H_{zz}+\left(q^2\dfrac{\ln ^{\prime}(c_T)c_T^2}{\ln ^{\prime}(c_X)c_X^2}-\omega ^2\right)H_1, \\
Z_{\phi }=\varphi +\dfrac{\phi ^{\prime }}{\ln ^{\prime }\left(c_X^{2(p-1)}\right)}\sum _aH_{aa}, \end{cases} \notag \\
&\quad (Z_{1})_{a}=qH_{ta}+\omega H_{za}, \notag \\
&\quad (Z_{2})_{ab}=H_{ab}-\delta _{ab}H_1, \label{ginvfluc}
\end{alignat} 
where $H_1=\frac{1}{p-1}\sum _a H_{aa}$. The fluctuations $(Z_0,Z_{\phi})$, $Z_1$ and $Z_2$ correspond to the three kinds of modes of \eqref{3channel}. Note that the number of modes
 \eqref{ginvfluc} is indeed equal to the number of degrees of freedom of gauge invariant fluctuations; $\frac{(p+2)(p+3)}{2}+1$  from graviton and dilaton minus $2(p+2)$ of diffeomorphism. 
    
 We use the following dimensionless ratios
 \begin{alignat}{4} 
&\mathtt{m}=\dfrac{\omega }{2\pi T},\quad \mathtt{q}=\dfrac{q}{2\pi T},  
\end{alignat} 
where $T$ is the Hawking temperature of the black $p$-brane \eqref{geometry},
\begin{alignat}{4} 
&T=\dfrac{7-p}{4\pi r_0}\left(\dfrac{r_0}{L} \right)^{\frac{(7-p)}{2}}  .
\end{alignat} 
Gauge invariant fluctuations \eqref{ginvfluc} follow linear equations of motion. From the analysis near the horizon by imposing the incoming boundary condition there, they are expressed as follows:
\begin{alignat}{4} 
&Z_A(r)=C_Af(r)^{-i\frac{\mathtt{m}}{2}}Y_A(r), \quad (A=0,\phi ,1,2) \label{bc2}
\end{alignat}
with $Y_A(r)$ analytic at $r=r_0$.  
Solving perturbatively for $Y_A(r)$ in the high temperature limit, $\mathtt{m}\ll 1$ and $\mathtt{q}\ll 1$, we impose the Dirichlet boundary
 conditions at infinity,
 \begin{alignat}{4} 
&Z_A(r)|_{r=\infty }=0. \label{bc}
\end{alignat}
 Boundary conditions ($\ref{bc}$) are important because these determine the dispersion relations. 
For the shear channel, we obtain
 \begin{alignat}{4} 
&(Z_1)_a(r)=(C_1)_af(r)^{-i\frac{\mathtt{m}}{2}}(1-f(r)+O(\mathtt{m},\mathtt{q}^2)), \label{quasi1}
\end{alignat}
and the familiar dispersion relation, 
\begin{alignat}{4} 
&\mathtt{m}=-i\dfrac{\mathtt{q}^2}{2}. \label{dispz1}
\end{alignat}
For the sound channel, we first evaluate the fluctuation\footnote{When we put the reflection of $\varphi$, $\varphi\to -\varphi$, under which $Z_{\phi}$ and $N_{\phi}$ interchange the roll: $N_{\phi}$ becomes invariant under the diffeomorphism. },
 \begin{align} N_{\phi}\equiv\varphi -\dfrac{\phi '}{\log '\left(c_X^{2(p-1)}\right) }\sum H_{aa}.\end{align} 
 $N_{\phi}$ satisfies the second order differencial equation, 
\begin{align}
N_{\phi}''+\log\left(\dfrac{c_Tc_X^p}{c_R}\right)N_{\phi}'+c_R^2\left(\dfrac{\omega ^2}{c_T^2}-\dfrac{q^2}{c_X^2} \right)N_{\phi}=0.\label{EE1}
\end{align}   
There are two independent solutions of \eqref{EE1} namely, the incoming wave type and the outgoing wave type solutions, and are no gauge dependent solutions which in general exist when we evaluate $H_{\mu\nu}$ in the next subsection. Thus, there is no diffeomorphisms which is consistent with \eqref{EE1} and change $N_{\phi}$.
Because of the invariance of $N_{\phi}$, we also impose the Dirichlet boundary condition \eqref{bc} on $N_{\phi}$, together with $Z_0$ and $Z_{\phi}$. We obtain
\begin{align}
&Z_0(r)=C_0f(r)^{-i\frac{\mathtt{m}}{2}}(1-f(r)+O(\mathtt{m}^2,\mathtt{q}^2)), \notag \\
&N_{\phi}=0,\label{soquasi}
\end{align}
and the dispersion relation,
\begin{alignat}{4} 
&\mathtt{m}=\sqrt{\dfrac{5-p}{9-p}}\mathtt{q}-i\dfrac{2}{9-p}\mathtt{q}^2+\cdot\cdot\cdot.\label{disp2}
\end{alignat}
$N_{\phi}$ is zero because there are no analytic solutions at $r=r_0$ except a constant, which we set to zero by the boundary condition \eqref{bc}. 
As $N_{\phi}=0$, $Z_{\phi}$ is proportional to $H_{1}$ for $p\neq 3$.  We show that for $p\neq 3$, 
$H_{1}|_{r=\infty}=0$ in the appendix A.3. Note that for $p=3$, $H_1$ is not invariant under the diffeomorphism, and $\varphi$ becomes independent of $H_{\mu\nu}$ (see the differential equations of $H_{\mu\nu}$ and $\varphi$ in the appendix A.2).

\subsection{Solutions for fluctuations $H_{\mu\nu}(r)$}

For the analysis based on the membrane paradigm, we have to evaluate Brown and York's boundary stress tensor near the horizon. However, the boundary stress tensor cannot be expressed solely by gauge invariant fluctuations; we need the functions $H_{\mu\nu}(r)$ themselves that compose gauge invariant fluctuations \eqref{ginvfluc}. In solving the equations of motion of $H_{\mu\nu}(r)$, we impose on $H_{\mu\nu}$ the incoming boundary condition near the horizon, up to the residual gauge transformations\footnote{From the analysis of the equations of motion of $H_{\mu\nu}$ near the horizon, we obtain three (four) independent non-singular solutions for the shear (sound) channel. In both channels, two of them are the incoming wave solution and the outgoing one. The other solutions correspond to the residual gauge transformations \cite{Policastro:2002tn}.}. 
 For each fluctuation, the Dirichlet boundary conditon at the infinity, $H_{\mu\nu}|_{r=\infty}=0$, may be imposed. This is the boundary condition adopted in the membrane paradigm approach \cite{Ko2}, but is not the one for the usual AdS/CFT (see also~\cite{Horo}). Using the Dirichlet boundary condition, we can read off viscosity of the dual field theory from constitutive relations of holographic stress tensor. 
      
For the shear channel, such functions $H_{\mu\nu}(r)$ which satisfy the incoming boundary condition up to the residual gauge
 transformation are
\begin{alignat}{4} 
&H_{ta}=\dfrac{1}{\mathtt{q}}\bigg\{ f(r)^{-i\frac{\mathtt{m}}{2}}\left(\dfrac{i\mathtt{q}^2}{2\mathtt{m}}f(r)+O(\mathtt{q^2})\right)+c\bigg\} , \notag \\
&H_{za}=\dfrac{1}{\mathtt{m}}\bigg\{ f(r)^{-i\frac{\mathtt{m}}{2}}\left(1-i\dfrac{\mathtt{m}}{2}+O(\mathtt{q}^2) \right)-c \bigg\} , 
\label{shsol}
\end{alignat}
where $c$ is an arbitrary constant corresponding to the freedom of residual gauge transformation, and 
 the factors $1/\mathtt{q}$ and $1/\mathtt{m}$ are multiplied for later convenience. 
 Of course, if these solutions \eqref{shsol} are substituted into the gauge invariant fluctuation $Z_1$ in \eqref{ginvfluc} and the Dirichlet boundary condition is imposed, we reobtain the quasi-normal mode \eqref{quasi1} and the dispersion relation \eqref{dispz1}. By imposing the Dirichlet boundary condition on the two fluctuations in \eqref{shsol}, we obtain 
$c=1$.

Next, let us consider $H_{\mu\nu}$ constituting the sound channel; $\chi=H_{zz}-H_1$, $H_{tz}$, $H_{tt}$ and $H_1$.
For the first two, we have
\begin{align}
 &\chi =H_{zz}-H_1=\dfrac{1}{\mathtt{m}^2}\bigg\{ f(r)^{-i\frac{\mathtt{m}}{2}}\left(-\dfrac{\mathtt{m}^2}{\mathtt{q}^2}p+\chi _q\right)+e\bigg\}, \notag \\
 &H_{tz}=\dfrac{1}{2\mathtt{qm}}\bigg\{f^{-i\frac{\mathtt{m}}{2}}(r)\left(-i\mathtt{m}(p-1)f(r)+O(\mathtt{q}^2)\right)-e+df(r)\bigg\},
 \label{sosol2}
\end{align}
where $e$ and $d$ are arbitrary constants of the residual gauge transformations, and $\chi _q$ is an $O(\mathtt{q})$ quatity, the precise form of which is not impotant below. If we impose the Dirichlet boundary conditions on $\chi$ and $H_{tz}$, we obtain $e=d=(\mathtt{m^2}/\mathtt{q^2})p+O(\mathtt{q})$.
 For $H_{tt}$ and $H_{1}$, we have not succeeded in constructing analytic solutions definded globally in all the region; we obtained solutions only near the horizon (and at the infinity). 
Taylor-expanding the analytic part at the horizon with respect to $x=r-r_0$, we obtain
\begin{alignat}{4} 
&H_{1}=\dfrac{1}{\mathtt{q}^2}f(r)^{-i\frac{\mathtt{m}}{2}}\left(1-\dfrac{(3-p)^2}{p}\dfrac{x}{r_0}+O(\mathtt{q},x^2)\right), \notag \\
&H_{tt} =\dfrac{1}{\mathtt{q}^2}\bigg\{f^{-i\frac{\mathtt{m}}{2}}(r)\left(-\dfrac{p-2}{3p}(3-p)^2\dfrac{x}{r_0}+O(\mathtt{q,x^2})\right)-d\bigg\}. 
\label{sosol}
\end{alignat}

\section{Holographic stress tensor}
In this section, we evaluate Brown and York's boundary stress tensor on the surface slightly away from the horizon to obtain the viscosity coefficients in the dual Yang-Mills theories. Though the equation of motion of the boundary stress tensor is given covariantly on the boundary of the black $p$-brane spacetime with fluctuations, we deform it so that it satisfies the conservation law in the flat space. Then, the dissipative parts of the deformed stress tensor can represent the energy momoentum tensor of the dual Yang-Mills theories.  
 At this time, we use the membrane paradigm. We consider the stretched horizon $\partial \mathcal{M}$ of dimension $p+1$ located at $r=r_h=r_0+\epsilon$ ($\epsilon\ll r_0$), and evaluate the deformed boundary stress tensor on $\partial \mathcal{M}$ which is composed of the gravity perturbations $H_{\mu\nu}$.
 
\subsection{Quasi-local stress tensor on black $p$-branes}
We consider the spacetime  $\mathcal{M}$ of the black $p$-brane \eqref{geometry} with fluctuation \eqref{f5} which has the stretched horizon $\partial \mathcal{M}$ located at $r=r_h$.
Let $g_{\mu\nu}$ and $n^{\mu}$ denote the metric of $\mathcal{M}$ (with fluctuations) and  the outward-pointing normal to $\partial \mathcal{M}$ with normalization $n^{\mu}n_{\mu}=1$. The induced metric on the stretched horizon,
 $\gamma _{\mu\nu}=g_{\mu\nu}-n_{\mu}n_{\nu}$, acts as a projection tensor onto $\partial \mathcal{M}$. The extrinsic curvature on $\partial \mathcal{M}$ is given by $\Theta   _{\mu\nu}=-\gamma _{\mu}^{ \text{ }\rho} \nabla _{\rho}n_{\nu}$. Following the analogy with the Hamilton-Jacobi approach (in the equation $H=-\frac{\partial S_{cl}}{\partial T}$, consider $T$ to be $\gamma_{\mu\nu}$), the stress tensor of Brown and York is given by 
 \begin{alignat}{4} 
&T ^{\mu\nu}\equiv \dfrac{2}{\sqrt{-\gamma }}\dfrac{\delta I}{\delta \gamma _{\mu\nu}}=\dfrac{1}{8\pi G_{p+2}}(\Theta ^{\mu\nu}-\gamma ^{\mu\nu}\Theta ^{ \text{ }\rho}_{\rho}). \label{BrownY}
\end{alignat}
In our case, the outward-pointing normal $n^{\mu}$ is 
\begin{alignat}{4} 
&\quad n^{\mu}=(0,\ldots ,0,1/c_R),
\end{alignat}
with $c_R$ given by \eqref{geometry}.
First, the fluctuations of the shear channel contribute only to the following three:
\begin{alignat}{4} 
&T^{z}{}_a=T^{a}{}_z=-\dfrac{n^r}{16\pi G_{p+2}}H_{az}'  , \notag \\
&T^t{} _{ a}=\dfrac{n^r}{16\pi G_{p+2}f}H_{ta}', \notag \\  
&T^{ a}{}_t=\dfrac{n^r}{16\pi G_{p+2}}\left(\dfrac{f'}{f}H_{at}-H_{at}'\right).
\label{shhol}
\end{alignat}
Note that there are no classical parts in these components of the boundary stress tensor.   
 Second, the components of stress tensor containing the fluctuation of the sound channel are
 \begin{alignat}{4} 
& T^{z}{}_z=\dfrac{n^r}{16\pi G_{p+2}}\left(\dfrac{f'}{f}+\dfrac{9-p}{r}-H_{tt}' +(p-1)H_{1}' \right), \notag \\
&T^{b}{}_a=\delta ^{b}{}_{a}\dfrac{n^r}{16\pi G_{p+2}}\left(\dfrac{f'}{f}+\dfrac{9-p}{r} -H_{tt}' +(p-2)H_{1}'+H_z' \right), \notag \\
&T^{t}{}_t=\dfrac{n^r}{16\pi G_{p+2}}\left(\dfrac{9-p}{r}+H_{zz}' +(p-1)H_{1}' \right), \notag \\
&T^t{}_{z} =\dfrac{n^r}{16\pi G_{p+2}f}H_{tz}', \notag \\
&T^{z}{}_t=\dfrac{n^r}{16\pi G_{p+2}}\left(\dfrac{f'}{f}H_{zt}-H_{zt}'\right),
\label{sohol}
\end{alignat}
where we have assumed that $H_{11}=H_{22}=...=H_1=\sum _{a=1}^{p-1}H_{aa}/(p-1)$.
Other components than \eqref{shhol} and \eqref{sohol} are of order $H^2$. 
\subsection{Conservation law}
For $g_{\mu\nu}$ and $\phi$ solving the equations of motion, the boundary stress tensor satisfies
 \begin{alignat}{4} 
&\mathcal{D}_iT ^i_{\ j} =-\tau _{\mu j}n^{\mu}=\dfrac{\beta}{16\pi G_{p+2}}\partial _j\phi\partial _r
\phi \, n^r
\label{conse}
\end{alignat}
where $\tau _{\mu\nu}$ is the matter stress-energy in the bulk, $\mathcal{D}_i$ is the covariant derivative on $\partial \mathcal{M}$, and the indices $i$ and $j$ run over the cordinates 
on $\partial \mathcal{M}$.  In both the shear channel and the sound channel, our solutions satisfy the equation ($\ref{conse}$) in the hydrodynamic limit. This equation is rewritten as
 \begin{alignat}{4} 
&\sqrt{-\gamma}\mathcal{D} _i w^i_{\ j}\equiv \partial _i(\sqrt{-\gamma}w^i_{\ j})-\sqrt{-\gamma}\Gamma ^k_{ij}w^i_{\ k}= 0, 
\label{conse2}
\end{alignat}
where $w^{i}{}_j$ is defined by
 \begin{equation}w^i{}_j=T ^i_{\ j}-\delta ^i_{\ j}\dfrac{\beta}{16\pi G_{p+2}}\varphi\partial _r\phi\, n^r. \label{T26} \end{equation}
 
For the shear channel, the last term of $(\ref{conse2})$ containing $\Gamma ^k_{ij}$ vanish. This means that 
$\sqrt{-\gamma}w^i{}_j$ satisfies the conservation law in the flat space, $\partial _i(\sqrt{-\gamma}w^i_{\ j})=0$.
Thus, $\sqrt{-\gamma}w^{i}_{\ j}$ can represent the conserved stress tensor of the dual field theory. One might be tempted to consider the symmetric tensor $\sqrt{-\gamma}w^{ij}$. However, this does not satisfy the conservation law in flat space. Moreover, there appear terms in $\sqrt{-\gamma}w^{ij}$ that contradict the constitutive relation of hydrodynamics.  

For the sound channel, $\sqrt{-\gamma}$ becomes $\sqrt{-\gamma _{(0)}}(-H_{tt}+H_{\alpha\alpha})/2$ where $\sqrt{-\gamma _{(0)}}$ has no fluctuations and $H_{\alpha\alpha}=H_{zz}+\sum _aH_{aa}$. To construct a quantity satisfying the conservation law in flat space for the sound channel, we expand the last term of $(\ref{conse2})$ with respect to the fluctuation: 

\begin{equation} 
\Gamma ^k_{ij}w^i_{\ k}=\dfrac{1}{2}(T_{(0) t}^tH_{tt},_j+T_{(0) 1}^1H_{\alpha\alpha},_j)+O(H^2), \label{K27} 
\end{equation}
where $T^i_{(0)j}$ is the stress tensor of the black $p$-brane
and it depends only on $r$. This implies that 
 \begin{equation} 
\mathcal{T}^i_{\  j}=\sqrt{-\gamma}(w^i_{\ j}-\delta ^i_{\ j}K),\label{T28}
\end{equation}
with $K$ given by
\begin{equation} 
K=\dfrac{1}{2}(T_{(0) t}^tH_{tt}+T_{(0) 1}^1H_{\alpha\alpha}), \label{K29}
\end{equation}
satisfies the conservation law in the flat space to $O(H)$:
\begin{equation}
\partial _i\mathcal{T}^i_{\  j}=0. 
\label{conse3}
\end{equation}
This $\mathcal{T}^i{}_j$ is regarded as the gravity counterpart of the energy momentum tensor of the dual Yang-Mills theories for the sound channel \cite{So}. Furthermore, since $\sqrt{-\gamma}w^i{}_j$ is equivalent to $\mathcal{T}^i{}_j$ for the shear channel (note that $\delta ^i{}_j=0$ for the shear channel), 
 we can take $\mathcal{T}^i{}_j$ for both the shear and sound channels.
 
\subsection{Hydrodynamics on the stretched horizon}
Our deformed boundary stress tensor $\mathcal{T}^{i}{}_j$ \eqref{T28} is related to the expectation value of the energy momentum tensor $\mathtt{T}^{ij}$ of the dual Lorentzian Yang-Mills theory in the following way:
 \begin{alignat}{4} 
&\eta _{jk}\vev{\mathtt{T}^{ik}}=\mathcal{T}^i{}_j, \label{A29}
\end{alignat}
where $\eta ^{ij}=\diag(-1,1,..,1)$ is the flat spacetime metric, and the right hand side should be evaluated on the stretched horizon $\partial \mathcal{M}$.\footnote{The Brown and York's boundary stress tensor at the infinity is related to the expectation value of the energy momentum tensor in the dual Yang-Mills theory by the relation \eqref{A29} (see \cite{My}). On the other hand, the boundary stress tensor on the stretched horizon is equal to the left hand side of \eqref{A29} up to the overall factor. This factor is not important for our later analysis.} The energy momentum tensor $\mathtt{T_{(P)}}^{ij}$ for the perfect fluid satisfies the constitutive relation:
\begin{alignat}{4} 
&\mathtt{T_{(P)}}^{ij}=(E +P)u ^iu ^j+P\eta ^{ij},
\label{C30}
\end{alignat}
where $E$, $P$ and $u ^i$ are the energy, the pressure and the fluid velocity of the dual Yang-Mills theory, respectively. Note that $i$ and $j$ run $p+1$ coordinates ($i,j=t,a,z$). The spatial part of $u ^i$ is equal to zero for the perfect fluid in the flat spacetime.
 
 To generalize the formula \eqref{C30} to the viscous fluid, one need to go to
the next order in the derivative expansion of the stress tensor, with at most one derivative in the spatial coordinates. The dissipative correction for the energy momentum tensor with the fluctuations of energy density $\delta E$ and that of pressure $\delta P$ is given by \cite{So}
\begin{alignat}{4} 
&\delta\mathtt{T}^{ij}=\delta E (u ^{i}u ^{j}+v_s^2 P^{ij}) -P ^i_{\ k}P ^j{}_{ l}\left(\zeta\eta^{kl}\partial _m u^m+ \eta
(\partial ^ku ^l+\partial ^lu ^k-\dfrac{2}{p}\eta^{kl}\partial _m u ^m )\right),
\label{constit}
\end{alignat}
 where $\eta$ and $\zeta$ are the shear and the bulk viscosity, respectively, $v_s=\sqrt{\delta P/\delta E}$ is the sound velocity, and 
$P^{ij}=\eta ^{ij}+u ^iu ^j$ is the projection tensor on the directions perpendicular to $u ^{i}$.

Let us rescale the spacial components $u^{\alpha}$ of the fluid velocity,   and the shear and the bulk viscosities by $E+P$ to express them in terms of
$\pi ^{\alpha}$, $\gamma _{\eta}$ and $\gamma _{\zeta}$, respectively:
 \begin{align}
 &u^{\alpha}=\dfrac{\pi ^{\alpha}}{E+P},
 \quad \eta =\gamma_{\eta}(E +P), \quad
\zeta=\gamma_{\zeta}(E +P).
 \end{align}
Then, the last term of $\delta \mathtt{T}^{ij}$ \eqref{constit} is expressed as follows:
\begin{alignat}{4} 
&-P ^i_{\ \alpha}P ^j{}_{ \beta}\left(\gamma_{\zeta}\eta^{\alpha\beta}\partial _\gamma  \pi ^{\gamma}+ \gamma_{\eta}
(\partial ^{\alpha }\pi ^{\beta}+\partial ^{\beta}\pi ^{\alpha}-\dfrac{2}{p}\eta^{\alpha\beta}\partial _\gamma \pi ^{\gamma} )\right) .
\label{T35}
\end{alignat}
Here, we have dropped the terms containing the derivatives of $u^t$ since $u^t=\sqrt{1+(u^{\alpha})^ 2}$ is equal to $1$ in the approximation of  keeping only terms linear in the fluctuation.

We now explain how to make the holographic identification \eqref{A29} of the above stress tensor $\mathtt{T}^{i}{}_{j}=\mathtt{T_{(P)}}^{i}{}_{j}+\delta\mathtt{T}^{i}{}_{j}$  with $\mathcal{T}^i{}_j$ on $\partial \mathcal{M}$. Especially, we identify the dissipative correction $\delta\mathtt{T}^{ij}$  with the $H_{\mu\nu}$ dependent terms in $\mathcal{T}^i{}_j$ (see \eqref{T26} and \eqref{T28} together with \eqref{shhol} and \eqref{sohol}).\footnote{The $H_{\mu\nu}$ independent terms of the boundary stress tensor  
 cannot reproduce the perfect fluid part  
of the dual Yang-Mills theory; energy density is zero compared with pressure because we calculate stress tensor not at the boundary but near the horizon.}  We have to check whether $\eta^{jk}\mathcal{T}^i{}_k$ is the symmetric tensor. 
  
For the shear channel, our identification \eqref{A29} becomes\footnote{Here, we consider the boundary stress tensor $\mathcal{T}^i{}_j$ on $\partial\mathcal{M}$ without any regularizations such as adopted in \cite{My}.} 
\begin{alignat}{4} 
&\mathcal{T}^t_{\ a}=\mathtt{T}^t{}_a=\pi _a,  \label{P36} \\
&\mathcal{T}^{z}{}_a=-\gamma _{\eta}(\partial _a\pi ^z+\partial ^z\pi _a)=-\gamma _{\eta}\partial ^z\pi _a, \label{P37}
\end{alignat}
where the $\partial _a\pi ^z$ term is dropped in the last expression of \eqref{P37} since we are considering fluctuations with a single fourier component 
$e^{-i(\omega t-qz)}$. Substituting \eqref{P36} for $\pi ^a$ into \eqref{P37}, using 
$(\ref{shhol})$ for $T^{i}{}_{j}$ and $(\ref{shsol})$ with $c=1$ (corresponding to the Dirichlet boundary condition at the infinity) for $H_{ta}$ and $H_{za}$, and evaluating \eqref{P37} on the stretched horizon ($r=r_h$)
, we obtain  
\begin{alignat}{4} 
&\gamma _{\eta}=\dfrac{1}{4\pi T}.
\label{shv}
\end{alignat}
This is the familiar result \cite{Policastro:2001yc} for the shear viscosity in the dual Yang-Mills theory for a black $p$-brane background. The same result is also obtained in \cite{Sa} using the membrane paradigm but in a different way.

We have to show that $\eta ^{jk}\mathcal{T}^i{}_k$ for the shear channel is a symmetric tensor, namely, that
\begin{align}
&\mathcal{T}^t{}_a=-\mathcal{T}^a{}_t, \label{T39} \\
&\mathcal{T}^z{}_a=\mathcal{T}^a{}_z. \label{T40}
\end{align}
The second equation follows from \eqref{shhol}.
The first equation is shown by imposing the Dirichlet boundary condition on $H_{ta}$ and $H_{za}$ and taking the limit $\mathtt{q}\to 0$ so that the condition $\mathtt{q}\ll -1/\log(\epsilon /r_0) $ is satisfied.

For the sound channel, we impose the Dirichlet boundary condition on the fluctuations $\chi$, $H_{tz}$, $Z_0$ and $N_{\phi}$. The Dirichlet boundary condition on
 $Z_{0}$ leads to the dispersion relation \eqref{disp2}.
The Dirichlet boundary condition for $\chi$ and $H_{tz}$  given by \eqref{sosol2} determines $e$ and $d$ as follows:
\begin{alignat}{4}
&e=\dfrac{\mathtt{m}^2}{\mathtt{q}^2}p-\chi _{q}, \notag \\
&d=\dfrac{\mathtt{m}^2}{\mathtt{q}^2}p-\chi _{q}+i\mathtt{m}(p-1),
\label{socons34}
\end{alignat}
 Let $\delta\mathcal{T}^i{}_j$ denote the linear term of 
 the fluctuation $H_{ij}$ in $\mathcal{T}^i{}_j$. Then, 
keeping only terms proportional to $1/\epsilon$, we have\footnote{In this calculation, we can neglect the last term of \eqref{T26}; it follows from \eqref{soquasi} that $\beta\varphi\phi '=\frac{1}{2r}(3-p)^2H_1$ with $H_1\sim\epsilon ^{-i\mathtt{m}/2}$. We also assume that $\epsilon/r_0\ll \mathtt{q}$ following \cite{Ko2}, and that $e^{-1/\mathtt{q}}\ll\epsilon /r_0 $ which is required by the symmetric tensor condition \eqref{T39}.   
Note that there appear terms containing $\mathtt{m}\log\epsilon$ which come from the series expansion of $f^{-i\mathtt{m}/2}$. However, such terms do not contribute to the hydrodynamics in our approximation.}  
\begin{alignat}{4}
&\delta \mathcal{T}^t{}_{t}=-\dfrac{e}{2\mathtt{m}^2}\dfrac{1}{\epsilon} \quad (=-\delta \mathtt{T}^{tt}=-\delta E), \notag \\
&\delta\mathcal{T}^t{}_{z}=\mathcal{T}^t{}_{z}=\dfrac{1}{2\mathtt{qm}\epsilon}\Bigl(d-i\mathtt{m}(p-1)(1+O(\mathtt{m}\log\epsilon )\Bigr) 
\quad (=\mathtt{T}^{tz}=\mathtt{\pi}^z), \notag \\
&\delta\mathcal{T}^z{}_{t}=\mathcal{T}^z{}_{t}=-\dfrac{e}{2\mathtt{qm}}\dfrac{1}{\epsilon}\quad (=-\mathtt{T}^{zt}), \notag \\
&\delta \mathcal{T}^z{}_{ z}=\dfrac{1}{2\mathtt{q}^2\epsilon}\Big(
d-i\mathtt{m}(p-1)(1+O(\mathtt{m}\log\epsilon ))\Bigr)\quad (=\delta \mathtt{T}^{zz}), \notag \\
&\delta  \mathcal{T}^a{}_{ b}=\delta ^a{}_b\dfrac{1}{2\mathtt{q}^2\epsilon}\Bigl(d
+i\mathtt{m}(1+O(\mathtt{m}\log\epsilon ))\Bigr)\quad (=\delta \mathtt{T}^{ab}),
\label{socons2}
\end{alignat}
 where the quantities inside the parentheses are the Yang-Mills counterparts, and we have omitted the common
factor $n^r\sqrt{-\gamma _{(0)}}/(16\pi G_{p+2})$ in the 
right hand sides. 
The origins of the right hand sides of \eqref{socons2} are two (recall that $\mathcal{T}^i{}_j$ is given by \eqref{T28} together with \eqref{T26}). One is $f'/f\sim 1/\epsilon$ in $T^z{}_z$ \eqref{sohol} multiplied by the terms in $\sqrt{-\gamma}$ linear in the fluctuation\footnote{ When we derive \eqref{socons2} and \eqref{F44}, we also use the fact that the order $\mathtt{q}$ terms of $H_{tt}$ and $H_{\alpha\alpha}$ are of the order $\epsilon ^{1-i\mathtt{m}/2}$;  $H_{\alpha\alpha}=(e+O(\epsilon ^{1-\mathtt{m}/2},\mathtt{q}\epsilon ^{1-\mathtt{m}/2}))/\mathtt{m}^2$ and $H_{tt}=(-d+O(\epsilon ^{1-i\mathtt{m}/2},\mathtt{q}\epsilon ^{1-\mathtt{m}/2}))/\mathtt{q}^2$. This follows by substituting \eqref{sosol2} and \eqref{sosol} into  \eqref{eomhtt} and \eqref{conshtt} in the appendix and evaluating $\chi_q$ in \eqref{sosol2} and the $O(\mathtt{q,x^2})$ terms in \eqref{sosol}.}:
\begin{alignat}{4}
&\sqrt{-\gamma}=\sqrt{-\gamma_{(0)}}(1-H_{tt}+H_{\alpha\alpha})=\sqrt{-\gamma_{(0)}}\left(1+\dfrac{e}{\mathtt{m}^2}+\dfrac{d}{\mathtt{q}^2}+O(\epsilon )\right).
\label{F44}
\end{alignat}
The other is the singular term $T^1_{(0)1}H_{\alpha\alpha}$ in \eqref{K29}.
The symmetric tensor condition for the sound channel, $\mathcal{T}^{t}{}_z=-\mathcal{T}^z{}_{t}$, is  satisfied on $\partial \mathcal{M}$ 
by taking the limit $\mathtt{q}\to 0$ as seen from \eqref{socons34} and \eqref{socons2}. 
We can calculate the shear viscosity $\gamma _{\eta}$ by considering the sound channel. For this, we use the
 identification $\mathcal{T}^{t}{}_z=\mathtt{T}^t{}_z=\pi _z$ and
\begin{alignat}{4}
&\delta\mathcal{T}^1{}_{1}-\delta\mathcal{T}^z{}_{z}=2\gamma _{\eta}\partial _z\pi _z.
\label{socons}
\end{alignat}
 From these and $(\ref{socons34})$ and $\eqref{socons2}$, we reobtain $(\ref{shv})$. Note that, on left hand side of \eqref{socons}, 
 the leading $\mathtt{1}/\mathtt{q}^2\epsilon$ terms cancel
  between $\delta\mathcal{T}^z{}_z$ and $\delta\mathcal{T}^1{}_1$ in \eqref{socons}.

Next, we calculate the sound velocity $v_s$ and the bulk viscosity $\gamma _{\zeta}$ of the dual Yang-Mills theory. 
We can obtain the sound velocity by substituting $\delta\mathcal{T}^z{}_z$ for $\delta\mathtt{T}^{zz}$ in the constitutive relation \eqref{constit}:
\begin{alignat}{4}
&\delta \mathcal{T}^z{}_{ z}=\delta  E\, v_s^2-\gamma_{\eta}\partial _z\pi _z\left(
\dfrac{\gamma_{\zeta}}{\gamma _{\eta}}+2-\dfrac{2}{p}\right).
\label{T43}
\end{alignat}
We can neglect the $\gamma_{\eta}\partial _z\pi _z$ term  on the right hand side of \eqref{T43}. This follows from the second and fourth equations of \eqref{socons2}: $\gamma _{\eta}\partial _z\pi _z/\delta\mathcal{T}^z{}_z=\gamma _{\eta}\partial _z\mathcal{T}^t{}_z/\delta\mathcal{T}^z{}_z\sim \mathtt{q}\ll 1$, where we have used the fact that  $\mathtt{q}$ and $\mathtt{m}$ are of the same order as seen from \eqref{disp2}.
Then, from \eqref{T43}, we obtain the sound velocity:
\begin{alignat}{4} 
&v_s^2=\dfrac{5-p}{9-p}.
\label{soundv}
\end{alignat}  
Furthermore, we can obtain the bulk viscosity $\gamma _{\zeta}$ by considering the subleading part of \eqref{T43}:  
\begin{equation} 
\gamma _{\zeta}=\dfrac{2(3-p)^2}{p(9-p)}\gamma _{\eta}. \label{B47}
\end{equation}

In fact, \eqref{soundv} and \eqref{B47} can be obtained from the conservation law \eqref{conse3}, the constitutive relation \eqref{C30} and \eqref{constit}, and 
the dispersion relation \eqref{disp2}. From \eqref{C30} and 
\eqref{constit}, we obtain \footnote{See also sec.\ 2.2 of \cite{So}} 
\begin{alignat}{4} 
&\omega =v_sk-i\dfrac{\gamma_{\eta}}{2}\left(\dfrac{\gamma _{\zeta}}{\gamma _{\eta}}+2-\dfrac{2}{p}\right) k^2.
\label{D48}
\end{alignat}
Comparing this with the dispersion relation \eqref{disp2} and using \eqref{shv}, we get \eqref{soundv} and \eqref{B47}. 
\section{Summary and discussions}
In this paper, we calculated the viscosity coefficients and the sound velocity of the dual Yang-Mills theory
using the membrane paradigm. 
This result is significant because we calculate the bulk viscosity of the dual theories by evaluating the coefficients of the boundary stress tensor not on the boundary at the infinity but on the stretched horizon $\partial\mathcal{M}$.  

The Dirichlet boundary condition at the infinity becomes important for our analysis because it determines the viscosity coefficients.
For the sound channel, we imposed the Dirichlet boundary conditions partially on $\chi$ , $H_{tz}$, $Z_0$ and $N_{\phi}$. For $p\neq 3$, $H_{tt}$ and $H_{11}$ also converge to zero at the infinity. 
The supergravity \eqref{ibulk} for $p=3$ is different from the supergravity for $p\neq 3$ because in the former, the fluctuation $\varphi$ doesn't interact with $H_{ij}$. 

We also want to explain the difference and the similarity between
the membrane paradigm and the usual AdS/CFT correspondence.
 In the sense of the gravity energy, the quasi-local stress tensor~\cite{Br,Ba} defined at the boudary of the AdS spacetimes can represent the expectation value of the energy momentum tensor of the dual theory in the strong coupling region~\cite{My}. In the same sense, we can 
 define the quasi-local stress tensor on $\partial\mathcal{M}$ as the gravity energy. However, as we explained in the section 3, it cannot show the perfect fluid part $\vev{\mathtt{T_{(P)}}^{ij}}$.
 As the quasi-local stress tensor $\mathcal{T}^i{}_j$ is defined on $\partial\mathcal{M}$, we guess that we cut the UV information near the boundary at the infinity which will contribute to $\vev{\mathtt{T_{(P)}}^{ij}}$. But in the hydrodynamic limit, (that means IR) we guess that the small deviation from the equilibrium may be seen. 
See also~\cite{BatoniAbdalla, Cantcheff:2007nr}. They explain how the strings dissipate at the horizons.  

\section*{Acknowledgements}
\hspace{0.51cm}
We would like to thank especially H. Hata for helpful comments to revise the  manusciript, and for valuable discussions.
We would like to thank T. Takayanagi and T. Azeyanagi for helpful discussions and comments. We also thank T. Takayanagi for the careful reading of the manuscript. 

\appendix

\section{Solving the equation of motion of the fluctuations}
In this appendix, we briefly summarize the derivation of 
\eqref{shsol} for the shear channel, and \eqref{sosol2} and \eqref{sosol} for the sound channel.
\subsection{Shear channel}

There are two second order differential equations
satisfied by $H_{ta}$ and $H_{za}$ and one constraint associated with
the gauge fixing condition $h_{\mu r}=0$:
 \begin{alignat}{4} 
&H_{ta}''+\ln '\left( \dfrac{c_X^{p+2}}{c_Tc_R}\right)H_{ta}'-q\dfrac{c^2_R}{c^2_X}(qH_{ta}+\omega H_{za})=0, \label{E49} \\
&H_{za}''+\ln '\left( \dfrac{c_Tc_X^{p}}{c_R}\right)H_{za}'+\omega\dfrac{c^2_R}{c^2_T}(qH_{ta}+\omega H_{za})=0, \label{E50} \\
&qH'_{za}+\omega \dfrac{c_X^2}{c_T^2}H_{ta}'=0.
\label{sheom}
\end{alignat}
These three equations are of course not independent; for example,
from \eqref{E49} and \eqref{sheom}, we obtain \eqref{E50}.
We multiply \eqref{E49} and \eqref{E50} by $r_0^2$, and \eqref{sheom} by $r_0/2\pi T$, 
and rewrite them in terms of dimensional quantites $\mathtt{q}$, $\mathtt{m}$ and $u=r_0/r$.
We also use the relation
\begin{align}
&T=\left.\dfrac{7-p}{4\pi r_0}\dfrac{c_X}{c_R\sqrt{f}}
\right|_{u=1}. 
\end{align}
Then, \eqref{E49}, \eqref{E50} and \eqref{sheom} are reduced to differential equations which have valuable $u$ and parameters $\mathtt{q}$ and $\mathtt{m}$, and are independent of $L$.
We consider the solution to them satisfying the incoming boundary condition,
\begin{alignat}{4} 
&H_{ta}=f^{-i\frac{\mathtt{m}}{2}}Y_{ta}(u), \\
&H_{za}=f^{-i\frac{\mathtt{m}}{2}}Y_{za}(u).
\end{alignat}
We consider the high temperature limit, and regard  $\mathtt{q}$ and $\mathtt{m}$ as being small and of the same order. Therefore, expanding the analytic parts as
\begin{align}
&Y_{ta}(u)=Y_{ta}^{(0)}\left(u,\dfrac{\mathtt{q}}{\mathtt{m}}\right)+\mathtt{q}Y_{ta}^{(1)}\left(u,\dfrac{\mathtt{q}}{\mathtt{m}}\right) +O\left(\mathtt{m}^2,\mathtt{q}^2,\mathtt{qm}\right), \label{S54}
\end{align}
and similary for $Y_{za}(u)$, we obtain the $c$-independent part of \eqref{shsol}. 
\subsection{Sound Channel}
We have to solve the following five second order differential equations and three constraints to derive \eqref{sosol2} and \eqref{sosol}:
\begin{alignat}{4} 
&H_{tt}''+\ln '\left( \dfrac{c_T^2c_X^{p}}{c_R}\right)H_{tt}'-\ln '(c_T)H'_{\alpha\alpha}-c^2_R\left(\dfrac{\omega ^2}{c^2_T}H_{\alpha\alpha}+\dfrac{q ^2}{c^2_X}H_{tt}+2\dfrac{q \omega}{c^2_T}H_{tz}\right)- \dfrac{2}{p}c_R^2\dfrac{\partial P}{\partial \phi}\varphi =0, \label{eomhtt} \\
&H_{tz}''+\ln '\left( \dfrac{c_X^{p+2}}{c_Tc_R}\right)H_{tz}'+\dfrac{c^2_R}{c^2_X}q\omega \sum _aH_{aa}=0, \label{E57} \\
&H_{zz}''+\ln '\left( \dfrac{c_Tc_X^{p+1}}{c_R}\right)H_{zz}'+\ln '(c_X)(H'_{aa}-H'_{tt})+c^2_R\left(\dfrac{\omega ^2}{c^2_T}H_{zz}+2\dfrac{q \omega}{c^2_T}H_{tz}+\dfrac{q^2}{c_X^2}(H_{tt}-\sum _aH_{aa})\right) \label{E58} \notag \\
&\qquad +\dfrac{2}{p}c_R^2\dfrac{\partial P}{\partial \phi}\varphi =0, \\
&H_{11}''+\ln '\left( \dfrac{c_Tc_X^{2p-1}}{c_R}\right)H_{11}'+\ln '(c_X)\left(H'_{zz}-H'_{tt}\right)+c^2_R\left(\dfrac{\omega ^2}{c^2_T}-\dfrac{q^2}{c^2_X}\right)H_{11}
+\dfrac{2}{p}c_R^2\dfrac{\partial P}{\partial \phi}\varphi =0, \label{eomh11} \\
&\varphi ''+\ln '\left( \dfrac{c_Tc_X^{p}}{c_R}\right)\varphi '+\dfrac{1}{2}\phi '(H'_{\alpha\alpha}-H'_{tt})+c^2_R\left(\dfrac{\omega ^2}{c^2_T}-\dfrac{q^2}{c^2_X}\right)\varphi -\dfrac{1}{\beta}c_R^2\dfrac{\partial ^2P}{\partial \phi ^2}\varphi =0,
\end{alignat}
and
\begin{alignat}{5}
&H '_{\alpha\alpha}+\ln '\left( \dfrac{c_X}{c_T}\right) H_{\alpha\alpha}+\dfrac{q}{\omega}H_{tz}'+2\dfrac{q}{\omega}\ln '\left( \dfrac{c_X}{c_T}\right)H_{tz}+\beta \phi '\varphi =0, \label{conshii} \\ 
&H '_{tt}-\ln '\left( \dfrac{c_X}{c_T}\right) H_{tt}+\dfrac{\omega}{q}\dfrac{c_X^2}{c_T^2}H_{tz}' -\sum _aH_{aa}' -\beta \phi '\varphi =0, \label{conshtt} \\ 
&\ln '(c_Tc_X^{p-1})H'_{\alpha\alpha}-\ln '(c_X^p)H'_{tt}+c^2_R\left(\dfrac{\omega ^2}{c^2_T}H_{\alpha\alpha}+2\dfrac{q \omega}{c^2_T}H_{tz}+\dfrac{q^2}{c_X^2}(H_{tt}-\sum _aH_{aa})\right) \label{soeom} \\
&\qquad -\beta \phi '\varphi '+c_R^2\dfrac{\partial P}{\partial \phi}\varphi =0, \notag
\end{alignat}
 where $H_{\alpha\alpha}=H_{zz}+\sum _aH_{aa}$. Among the eight equations given above, we consider four equations, 
 \eqref{E57}, \eqref{E58} subtracted by \eqref{eomh11}, \eqref{conshii} and \eqref{conshtt}, for obtaining four unknown functions $H_{tz}$, $\chi =H_{zz}-H_{1}$, $H_1=\sum _a H_{aa}/(p-1)$ and $H_{tt}$. Recall that $\varphi$ is given in terms of $H_{1}$ (see footnote 6). The remaining four equations can be derived from the four equations we solve. We rewrite the four equations in terms of $u$, $\mathtt{q}$ and $\mathtt{m}$ and consider  the solutions satisfying the incoming boundary condition at the horizon, 
$H_{ij}=f^{-i\mathtt{m}/2}Y_{ij}$ and $\chi=f^{-i\mathtt{m}/2}Y_{\chi}$, as in the case of the shear channel.  
We can obtain $Y_{tz}^{(0)}$ and $Y_{tz}^{(1)}$ from \eqref{E57}, and $Y_{\chi}^{(0)}$ 
from \eqref{E58} subtracted by \eqref{eomh11}.\footnote{We can determine $Y_{tz}^{(0)}$, $Y_{tz}^{(1)}$ (up to the overall factor) and $Y_{\chi}^{(0)}$ without considering the $O(\mathtt{q}^2)$ terms in \eqref{E57} and \eqref{E58} subtracted by \eqref{eomh11}. The overall factor of $Y_{tz}^{(1)}$ can be fixed by evaluating the $1/(u-1)$ terms in the $O(\mathtt{q^2})$ part of \eqref{E57}. }  
In solving \eqref{conshii} and \eqref{conshtt} for $H_{tt}$ and $H_{1}$, we Taylor-expand $f$ and $Y_{tt}$ and $Y_{1}$ with respect to $u-1$ in each order of $\mathtt{q}$ and determine the coefficients. 
\subsection{The Dirichlet boundary condition in the sound channel}
The sound channel fluctuations \eqref{sosol2} and \eqref{sosol} are obtained by adding the gauge terms ($e$ and $d$ terms)
to the solutions given in A.2. In sec. 3.3, we imposed
 the Dirichlet boundary condition $H_{ij}|_{u=\infty}$ on $\chi$ and $H_{tz}$
 to fix the values of $e$ and $d$ as given by \eqref{socons34}.  For $p\neq 3$, the solutions $H_{tt}$ and $H_1$ in ($\ref{sosol}$) also satisfy the same Dirichlet boundary condition at the infinity. To see this, we consider the behavior of the solution \eqref{sosol} at the infinity.  We expand $H_{1}$ in powers of $1/u$, 
\begin{align}
H_{1}=\sum_{k=n}^{\infty} c_{1}^k\dfrac{1}{u^k} \quad (n \ \text{is integer}), 
\end{align}
and substituting this into \eqref{conshii} and using the explicit formula of \eqref{sosol2}, we find $c_{1}^l=0$ $(l\le 0)$.
Then, from the comparision with the quasi-normal modes $(\ref{soquasi})$ which is composed of $H_{ij}$,
\begin{align}
Z_0|_{u=\infty}=\left. q^2\dfrac{c_{T}^2}{c_X^2}H_{tt}+2q\omega H_{tz}+\omega ^2H_{zz}+(q^2\dfrac{\ln ^{\prime}(c_T)c_T^2}{\ln ^{\prime}(c_X)c_X^2}-\omega ^2)H_1\right|_{u=\infty} =0,
\end{align}
 we can see that $H_{tt}|_{u=\infty}$ 
and the others go to zero at the infinity with the gauge dependent terms given in \eqref{socons34}.

 \end{document}